# Agreement Is Not Quality: Blind Expert Verification of Human and LLM Qualitative Coding When Human Consensus Is Not Ground Truth

Alex Liu, University of Washington. alexliux@uw.edu (Corresponding Author)
Lief Esbenshade, University of Washington. lief@uw.edu
Michael Xiao, University of Washington. mxiao16@uw.edu
Victor Tian, University of Washington. ztian27@uw.edu
Zachary Zhang, Colleague AI/ zac@colleague.ai
Kevin He, Colleague AI. kevin@colleague.ai
Min Sun, University of Washington. misun@uw.edu

**Abstract.** Evaluations of LLM-assisted qualitative coding almost universally measure model performance as agreement with human coders, a practice that presumes human coding is the standard to approximate. This study provides empirical evidence that the presumption fails in ways agreement metrics cannot detect. Five LLM systems and three trained human coders independently applied a 72-item hierarchical codebook to 2,560 educator messages from a K-12 AI platform. Beyond conventional agreement analysis, an independent domain expert judged 855 pairwise comparisons of code sets blind to source, treating human and machine sources symmetrically. The two evaluation approaches diverge in both directions. Human-LLM agreement (mean Jaccard 0.30) falls well below human-human agreement (0.52), which standard practice would read as inferior LLM coding, yet the blind verifier preferred human and LLM coding at indistinguishable rates (51.5% vs. 48.5%, $p = 0.537$), and a Bradley-Terry ranking placed two LLMs above two of three human coders. For several substantive codes, human consensus encoded shared bias that the verifier rejected in favor of the LLM interpretation. Agreement-based evaluation is therefore insufficient for automation decisions, and the study demonstrates a transferable verification protocol and a code-level division-of-labor framework.

*Keywords*: Large language models (LLMs), computer-based qualitative analysis, human-AI collaboration, text annotation, text as data, evaluation methodology

## Introduction

Social science research is currently making consequential decisions about whether qualitative coding can be delegated to large language models (LLMs), and the debate has sharpened into two poles. On one side, systematic evaluations report that LLMs can outperform expert human coders and supervised classifiers (Törnberg, 2025). On the other, methodological work warns that LLM annotation carries epistemic risks across validity, reliability, replicability, and transparency that performance metrics do not capture (Lin & Zhang, 2025; Baumann et al., 2025). Both positions nonetheless share an evidentiary base, since nearly all evaluations on either side rest on a single form of evidence, namely how closely LLM output agrees with the output of trained human coders (e.g., Ashwin et al., 2025; Misiejuk et al., 2025). This validation logic treats human coding as the standard to approximate. The logic is compelling when human coders converge on a shared and valid interpretation, but its limitations become more apparent in contexts where interpretations are inherently uncertain or contested. Complex, comprehensive coding instruments applied to large naturalistic datasets are the settings in which researchers most need scalable coding assistance, and they are also the settings in which trained human coders themselves achieve only moderate agreement with one another (Hennessy et al., 2020). When humans disagree, deviation from any one human coder cannot distinguish an LLM that codes poorly from an LLM that codes differently but well. When humans agree, their consensus may still reflect shared interpretive bias rather than validity, a possibility that agreement-based evaluation is structurally unable to detect.

The present study provides direct empirical evidence on both possibilities. It does so by subjecting human and LLM coding to the same blind evaluation. Five LLMs and three trained human coders independently applied a 72-item hierarchical codebook to 2,560 unstructured educator messages, and an independent domain expert then judged 855 pairwise comparisons of their code sets without knowing which source produced which set. Because the protocol treats human and machine sources symmetrically, it can answer a question that no agreement study can ask. When two sources disagree, which interpretation does an independent expert prefer, and does that preference track the pattern that agreement metrics would predict?

The answer, developed across the findings, is that it does not. Agreement metrics and expert quality judgments diverge in both directions. Human coders converge on interpretations that the blind verifier rejects in favor of the LLM reading, indicating that human consensus can encode shared conservative bias. Moderate human-LLM agreement coincides, for other codes, with low expert endorsement of both sources, indicating that agreement can certify shared error. These two failure modes are invisible to the evaluation designs used across the existing literature, whether those designs reach optimistic conclusions (Chew et al., 2023; Dunivin, 2025) or cautionary ones (Ashwin et al., 2025; Baumann et al., 2025), because both camps measure quality within the agreement paradigm.

The study builds on conceptual work calling for this shift. Thomas et al. (2025) proposed moving from agreement-as-quality to disagreement-as-information in educational AI annotation but did not offer operationalizable guidelines, and no prior study has placed human coders under the same anonymized scrutiny as the models they are used to validate (Zheng et al., 2023; Norman et al., 2026). That design move is the study's central methodological contribution, and it is what makes the divergence findings observable.

The evaluation setting is deliberately demanding. The study uses a 72-code hierarchical, multi-label instrument developed through prior inductive analysis, applied to unstructured educator messages spanning any instructional topic. This scale is representative of comprehensive qualitative instruments in practice, and it produces the condition the study is designed to examine, namely genuine interpretive difficulty. This setting allows us to investigate the human-AI collaborative dynamics that prior literature suggests are advantageous for qualitative analysis (Liu & Sun, 2025; Liu et al., 2025).

### Research Questions

Four questions structure the study. The first two establish the agreement picture that standard evaluation practice would produce, the third subjects that picture to the blind verification test, and the fourth translates the combined evidence into guidance for practice.

**RQ1.** To what extent do LLMs agree with trained human coders when applying a complex multi-label codebook to unstructured educational data, and how does this agreement compare to human-human agreement?

**RQ2.** Under what conditions does human-LLM alignment increase or decrease, and what code-level characteristics predict where models succeed or struggle?

**RQ3.** When human and LLM coders disagree, which source does an independent domain expert prefer, and how do all sources rank relative to one another in a symmetric evaluation?

**RQ4.** What practical guidelines emerge for determining the division of labor between human coders and LLMs on complex qualitative instruments?

## Related Work

### LLM-Assisted Qualitative Coding and Its Moderators

Evidence that LLMs can perform deductive coding has accumulated rapidly across the social sciences. Chew et al. (2023) showed that GPT-3.5 reached agreement comparable to human coders across four benchmark datasets, and Dunivin (2025) reported kappa of 0.60 or above for eight of nine socio-historical codes using GPT-4 with reasoning, establishing that large-scale content analysis is feasible under appropriate conditions. Törnberg (2025) reported the strongest result in this direction, finding that instruction-tuned models outperformed both expert coders and supervised classifiers at annotating political social media messages. In interpretive qualitative work, De Paoli (2024) demonstrated that an LLM can perform inductive thematic analysis of semi-structured interviews while documenting the limits of the approach. Subsequent work located the boundary conditions. Liu et al. (2025) found that construct characteristics such as clarity, concreteness, and objectivity predicted GPT-4 performance better than codebook size, and Than et al. (2025) showed that instruction-tuned models can replicate iterative coding on long documents with substantial variation across code types. Ashwin et al. (2025) demonstrated that LLMs can introduce serious bias, with over-prediction ratios exceeding five to one on certain categories. Orchestration studies complicate the picture further. Ahtisham et al. (2026) found that self- and cross-verification improved Cohen's kappa by 58% over an unverified baseline, while Borchers et al. (2025), testing six LLMs in 18 configurations across more than 77,000 coding decisions, found that single agents matched or outperformed multi-agent consensus.

Codebook complexity is among the strongest moderators in this literature. Binary or small-category schemes generally yield high agreement (Chew et al., 2023), moderate codebooks of 10 to 30 codes yield widely varying results depending on code properties (Liu et al., 2025), and even trained human coders struggle to apply complex schemes reliably (Hennessy et al., 2020). Large hierarchical label spaces have been studied extensively in adjacent text classification research, but those evaluations treat coding as benchmark classification against fixed labels rather than as interpretive qualitative analysis. Within the qualitative coding literature, evaluations on instruments approaching the complexity of comprehensive codebooks used in practice remain rare, and existing work at this scale has relied on agreement-based evaluation. The present study operates on a naturally large instrument (72 codes, hierarchical, multi-label) that emerged from inductive analysis rather than experimental construction, and subjects it to an evaluation design that does not presuppose a gold standard.

**Evaluation Without Assuming a Gold Standard**

How to evaluate coding quality when neither source constitutes ground truth has moved from theoretical concern to active methodological development. The concern has a longer lineage in content analysis methodology. Campbell et al. (2013) showed that intercoder reliability on in-depth interview data must be actively constructed through unitization and iterative scheme refinement rather than assumed, and Nelson et al. (2021) benchmarked computational text analysis against rigorous hand-coding while explicitly noting that hand-coding serves as the reference by convention rather than by demonstrated validity. Lin and Zhang (2025) brought this concern to LLM annotation, showing that performance degrades as task complexity increases, with multilabel classification notably weaker than binary tasks, and arguing that validity, reliability, replicability, and transparency require separate assessment. Thomas et al. (2025) proposed reconceptualizing ground truth for educational AI annotation through four shifts, including from agreement-as-quality to disagreement-as-information and from source-blind to source-aware evaluation. Norman et al. (2026) provided large-scale evidence that LLM judges can exhibit reliability without validity, and Baumann et al. (2025), testing 18 LLMs across 21 datasets and over 13 million annotations, showed that models can produce consistent but biased outputs that standard metrics score as reliable. Converging applied findings point the same way. Misra et al. (2026) found that open-source LLMs approximate human gold-standard codes inconsistently across topic areas, and Wen et al. (2026) found that LLM-supported thematic analysis requires researcher oversight for contextually interpretive codes.

Preference-based evaluation offers an alternative to agreement against a fixed reference. Zheng et al. (2023) demonstrated through MT-Bench and Chatbot Arena that pairwise preference judgments, modeled in the Bradley-Terry framework, can estimate relative quality without a gold standard. The present study adapts this approach to qualitative coding, using a domain expert rather than a language model as the judge to preserve domain authority.

The reviewed literature converges on a shared structure. Whether studies reach optimistic conclusions (Törnberg, 2025; Chew et al., 2023; Dunivin, 2025) or cautionary ones (Lin & Zhang, 2025; Baumann et al., 2025; Ashwin et al., 2025), they measure quality within the agreement paradigm, treating human coding as the reference and model output as the object of evaluation. Conceptual arguments for moving beyond this paradigm exist (Thomas et al., 2025), as do warnings that agreement metrics overstate reliability (Baumann et al., 2025; Norman et al., 2026). Therefore, we propose a study that implements an evaluation in which human and LLM sources are judged blind, symmetrically, and by independent domain expertise. The present study's design makes it possible to test, rather than assume, whether agreement corresponds to quality, under conditions representative of the settings where decisions about the division of labor between humans and LLMs arise.

# Methodology

## Study Context and Data

This study evaluates human and LLM coding using a corpus of educator-AI conversations from a [redacted generative AI platform name] for K-12 instructional support serving educators. The randomly selected dataset comprises 2,560 educator-AI messages spanning K-12 grade levels, subject areas from literacy to career and technical education, and diverse instructional contexts, covering 500 conversations occurred in the period from June 2025 to June 2026. The current study is part of a larger research project, in which the codebook was developed through prior inductive analysis of a different set of sampled conversations from the same source. All data were collected under the platform's terms of use, personally identifiable information was removed prior to analysis, and the study protocol was reviewed and approved by the [redacted institution name]'s Institutional Review Board. All LLM coding was conducted through enterprise API accounts with contractual terms confirming that submitted data would not be used for model training. Each message was coded through a separate API call, which ensured the independence of the coding results.

**Codebook**

The coding instrument is a hierarchical codebook, developed to conceptualize how K-12 educators use AI, comprising 72 instructional items organized within 19 mid-level categories and 6 top-level domains. The codebook was developed through a multi-phase inductive process involving open coding, axial coding, and selective coding, followed by iterative refinement through code frequency analysis, semantic distinctiveness evaluation, and alignment with pedagogical literature.

The six domains are Instructional Practices (24 items), Curriculum and Content Focus (16 items), Student Needs and Context (12 items), Assessment and Feedback (10 items), Professional Responsibilities (6 items), and Other (4 items); the full codebook is included in Appendix E. Coding is multi-label, meaning each message may receive zero or more codes from instructional items. In practice, educator messages receive an average of 2 codes.

Several properties make this codebook demanding for both human and machine coders. Codes are not applied as independent decisions, since assigning one can depend on how others are interpreted. Some codes are identifiable from surface language, while others require inferring pedagogical intent that is not explicitly stated.

## Human Coding

Three human coders independently coded the full corpus. The team comprised researchers with a doctoral degree in education, a Master's degree in education, and an undergraduate minor in education with industry research experience, together contributing expertise in instructional design, curriculum, pedagogy, and applied research methodology. Prior to independent coding, the team completed three rounds of calibration following established practices (Hennessy et al., 2020). During calibration, the human coders discussed the codebook and coding results to align their understanding and recorded coding decisions and borderline cases. Through this process they collectively developed a coding protocol that guided their independent work. Operational agreement on the designated overlap set (approximately 285 messages from 50 conversations coded by all three coders) reached a Jaccard similarity of 0.52, consistent with published benchmarks for complex multi-label coding and reflecting interpretive variability inherent to the instrument rather than coder deficiency. Human coders also authored the item descriptions included in the codebook for LLM coding, reflecting their collective understanding of each code.

## LLM Coding

Five LLMs independently coded the same corpus and entered the verification protocol. These focal models were drawn from a larger pool that coded the corpus under identical conditions, and were selected to cover the major commercial providers, span capability tiers, and include widely adopted systems. Because verifier capacity was limited, models outside the focal set did not enter verification and are reported in Appendix A for completeness.

**Table 1.**

Five Focal Models

| Model | Provider | Tier |
|---|---|---|
| Claude Opus 4.8 (Claude Opus) | Anthropic | Large-scale reasoning model |
| Claude Haiku 4.5 (Claude Haiku) | Anthropic | Lightweight model |
| Gemini 3.5 Flash (Gemini Flash) | Google | Fast inference model |
| GPT-4o | OpenAI | Multi-modal flagship model |
| GPT-5.5 | OpenAI | Latest-generation model |

*Note. Gemma 4 (Google), GPT-5.4 Mini (OpenAI), and GPT-5.4 Nano (OpenAI) coded the corpus under identical conditions to the five focal models but were not included in the blind verification protocol due to verifier capacity constraints. Full API model strings and access dates, where available, appear in Appendix C for all models.*

To ensure parity, every system was supplied with an identical chain-of-thought structured prompt. The prompt included the complete codebook with human coders-authored item definitions, explicit directives to identify all relevant codes, and a specified JSON response format. Full prompt, model identifiers, configuration parameters, and access dates appear in Appendix C.

To maximize determinism, models were configured with temperature zero or the minimal thinking capacity the provider permits, no session memory, and messages submitted individually. Each model coded the corpus excluding 66 platform trigger phrases (n = 2,494). To assess stability, four of the focal models were run a second time under identical conditions; intra-model exact match ranged from 71.1% to 99.2% and Cohen's kappa from 0.871 to 0.995, exceeding the human-human baseline and confirming that human-LLM disagreements reflect systematic interpretive differences rather than stochastic noise.

**Analysis Framework**

Given the multi-label structure and the absence of a presumed gold standard, the analysis employs complementary agreement measures at multiple levels of granularity.

The primary measure is Jaccard similarity, computed for each message between two code sets as J(A, B) = |A intersection B| / |A union B|. Jaccard ranges from 0 (no overlap) to 1 (identical sets), handles varying set sizes naturally, and is not inflated by shared absences. With 72 possible codes and an average of 2 assigned per message, approximately 97% of codes are correctly not assigned at any given time. Metrics crediting shared absences would be dominated by trivial agreement.

Directional hit rates capture asymmetric patterns. Hit(H to L) measures the proportion of human-assigned codes also appearing in the LLM annotation (recall). Hit(L to H) measures the proportion of LLM-assigned codes also appearing in the human annotation (precision). Divergence between these rates reveals systematic over-coding or under-coding tendencies.

Multi-model consensus measures how many models agree on a given code for a given message and relates this to the rate of human endorsement.

**Blind Verification Protocol**

To evaluate coding quality without presupposing either source's superiority, we designed a blind verification protocol adapted from preference-based evaluation paradigms (Zheng et al., 2023), drawing on paired comparison methods formalized by Bradley and Terry (1952), in which relative quality is estimated from pairwise preferences rather than comparison against a fixed standard. Figure 1 summarizes the design.

**Figure 1**.

*Study design.*

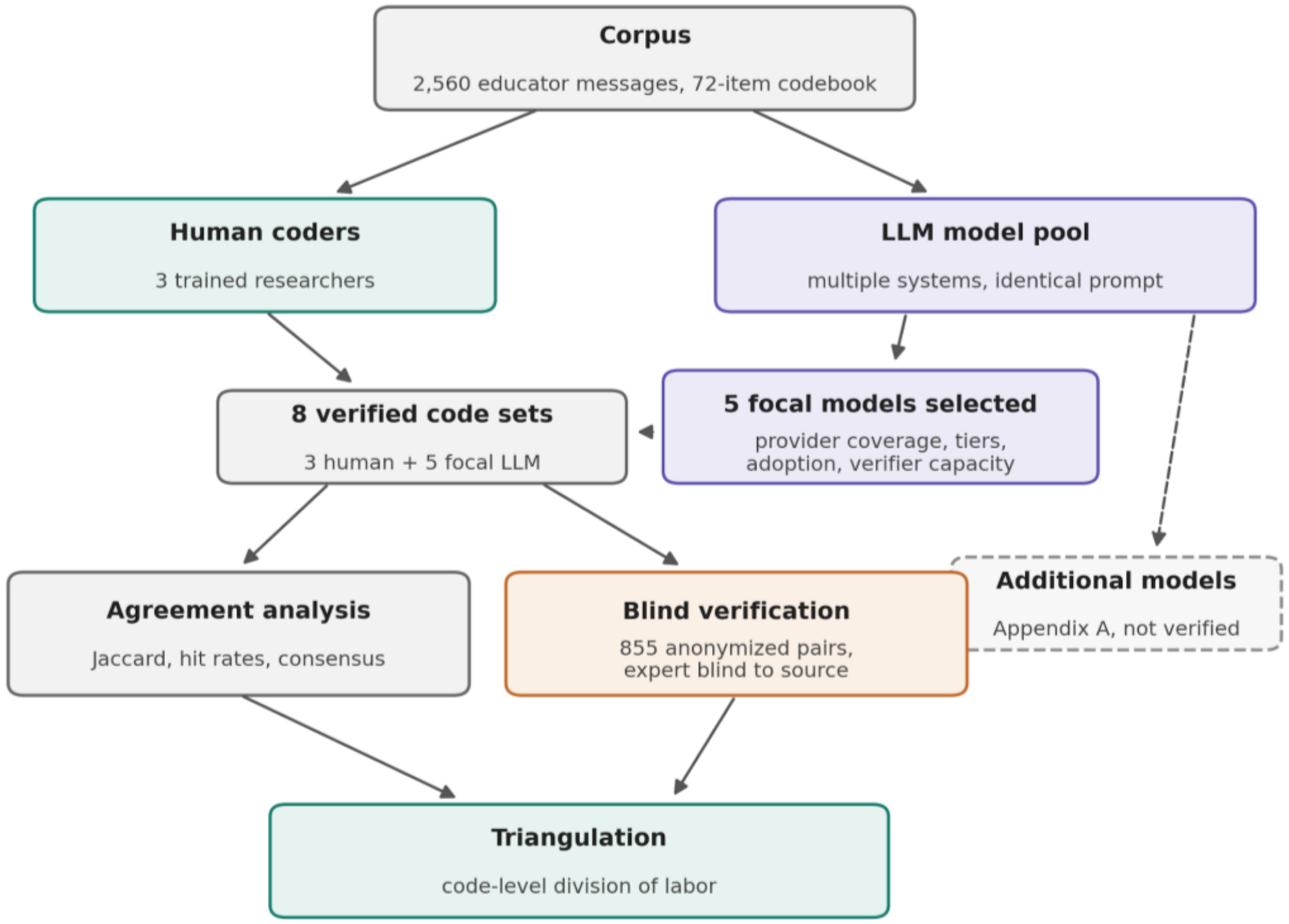

*Note. A corpus of 2,560 educator messages is independently coded by three trained human coders and by a pool of LLM systems, from which five focal models enter verification on the basis of provider coverage, capability tiers, adoption, and verifier capacity; remaining models are reported in Appendix A. Outputs feed two parallel evaluation tracks, a conventional agreement analysis and a blind verification protocol in which an independent domain expert judges 855 anonymized pairwise comparisons. Triangulating the two tracks produces the code-level division-of-labor classification.*

The verifier is a researcher with a doctoral degree in education, extensive educational research experience, and K-12 teaching experience, who was not involved in the original coding and had no knowledge of which sources produced which annotations. A stratified random sample of 855 message-level comparison pairs was constructed. The sample size reflects the practical capacity of a single expert verifier, and verification was limited to the five focal models so that each comparison stratum retained enough pairs for meaningful analysis. Each pair presents two code sets for the same message without revealing source identity, with position randomized, and the verifier selects one of four responses: Set 1 preferred, Set 2 preferred, Both adequate, or Neither adequate.

The pairs comprise 555 Human vs. LLM pairs (primary analysis), 210 LLM vs. LLM pairs (model ranking), 85 Human vs. Human pairs (baseline calibration), and 5 code supplement pairs for rare codes. Results support a binomial test of preference direction, a Bradley-Terry model over the eight verified sources, per-code endorsement rates, and tie rates.

## Findings

Sections 4.1 and 4.2 report the picture that agreement-based evaluation would produce on its own. The blind verification findings (RQ3) subject that picture to the verification test, and the division-of-labor analysis (RQ4) translates the combined evidence into practice.

### RQ1: Overall Human-LLM Agreement

Table 2 presents mean Jaccard similarity between each LLM and the three human coders across the 2,210 eligible messages.

**Table 2.**

*Human-LLM Agreement by Model.*

| Model | Mean Jaccard | Hit(H to L) | Hit(L to H) | Exact Match % |
|---|---|---|---|---|
| **Gemini Flash** | 0.340 | 47.0% | 47.1% | 17.6% |
| **GPT-5.5** | 0.330 | 48.4% | 44.8% | 16.7% |
| **Claude Opus** | 0.322 | 50.0% | 42.1% | 14.0% |
| **Claude Haiku** | 0.276 | 42.0% | 40.3% | 13.5% |
| **GPT-4o** | 0.234 | 27.0% | 50.1% | 14.3% |

*Note. Mean Jaccard similarity, directional hit rates, and exact match rates between each LLM and the three human coders, computed across 2,210 eligible messages. Coding results for additional models are in Appendix A.*

Mean human-LLM Jaccard across models was 0.30 (range 0.23 to 0.34), with exact match rates of 13.5% to 17.6%. On the overlap set of 284 messages coded by multiple humans, pairwise human-human Jaccard mean approximately 0.52, leaving a gap of roughly 0.22 Jaccard points between human-human and human-LLM agreement. Pairwise agreement among the five LLMs ranged from 0.37 to 0.68, with Gemini Flash, GPT-5.5, and Claude Opus forming a high-agreement cluster and GPT-4o the most divergent model. LLM-LLM agreement is therefore comparable to human-human agreement and notably higher than human-LLM agreement, indicating that models share interpretive tendencies that differ from human patterns. All three coders exhibited the same relative ordering of model agreement, so the human-LLM gap reflects a structural difference rather than idiosyncratic coder variation.

## RQ2: Conditions Moderating Agreement

### *Code-Level Stratification*

The 72 codes separate into three tiers by mean human-LLM Jaccard. Tier 1 codes (above 0.4, 8 codes) are marked by concrete task language identifiable from surface cues, such as Grading (0.63), Generate Feedback to Students (0.57), and Learning Standards Alignment (0.46). Tier 2 codes (0.2 to 0.4, approximately 30 codes) require some inference but retain recognizable textual patterns, such as ELA Skills Development and Unit Planning; at model confidence of 0.9 or above, hit rates for these codes reach 57% to 75%. Tier 3 codes (below 0.2, approximately 35 codes) require inferring intent or recognizing pedagogical patterns that models resolve differently from humans, including high-frequency codes such as Non-Educational (Jaccard 0.05) and Professional Development Needs (0.03).

### *Directional Patterns*

Gemini Flash, GPT-5.5, and Claude Opus exhibit balanced profiles, with hit rates within five percentage points in both directions and coding volumes comparable to humans (1.64 to 2.10 codes per message vs. the human average of 1.52). Claude Haiku shows lower alignment in both directions. GPT-4o exhibits a strongly conservative profile, with high precision (50.1% of its codes endorsed by humans) but low recall (capturing 27.0% of human-assigned codes), and leaves 40.7% of messages entirely uncoded compared to 2.2% for human coders, suggesting a qualitatively different coding strategy rather than merely weaker performance.

### *Multi-Model Consensus*

The relationship between the number of models agreeing on a code and the rate of human endorsement is monotonically increasing.

**Table 3.**

*Human Endorsement by Level of Multi-Model Consensus*

| **Models agreeing** | **Human also assigns** |
|---|---|
| 1 of 5 | 12.2% |
| 2 of 5 | 22.4% |
| 3 of 5 | 31.4% |
| 4 of 5 | 48.5% |
| 5 of 5 | 63.0% |

*Note. For codes assigned by a given number of the five LLMs, the percentage that human coders also assigned. Endorsement rises monotonically with model agreement, from 12.2% when a single model assigns a code to 63.0% when all five converge.*

When all five models converge on a code, human coders also assign it 63% of the time. Conversely, 32% of human-assigned codes are assigned by no model, representing interpretations that appear to require human expertise.

*Confidence as a Quality Signal*

Model-reported confidence provides a quality signal whose informativeness varies across models, and characterizing it requires identical confidence bins for every model. Pooled, agreement rises with confidence (66% at 0.9 or above vs. 29% below 0.7; point-biserial $r = 0.197$, $p < 0.001$), but model-level patterns differ in shape. Claude Opus rises monotonically from 17% in its lowest bin to 83% at 0.9 or above, a lift of 54 points between pooled low and high confidence. GPT-5.5 and Claude Haiku show moderate lifts of 37 and 43 points, though Haiku is non-monotonic at low confidence. GPT-4o never reports confidence above 0.9; taking its highest emitted bin (0.8 to 0.9) as the ceiling, agreement rises from 27% to 64%. Gemini

Flash rarely reports low confidence (under 9% of annotations fall below 0.7) and is nearly flat between 0.5 and 0.9, discriminating only in its top bin. Triage is therefore viable only for models whose confidence is calibrated on the target instrument, with thresholds respecting each model's emitted range. Figure 2 visualizes these differences.

**Figure 2**.

*Human-Agreement Rate by Model-Reported Confidence Bin*

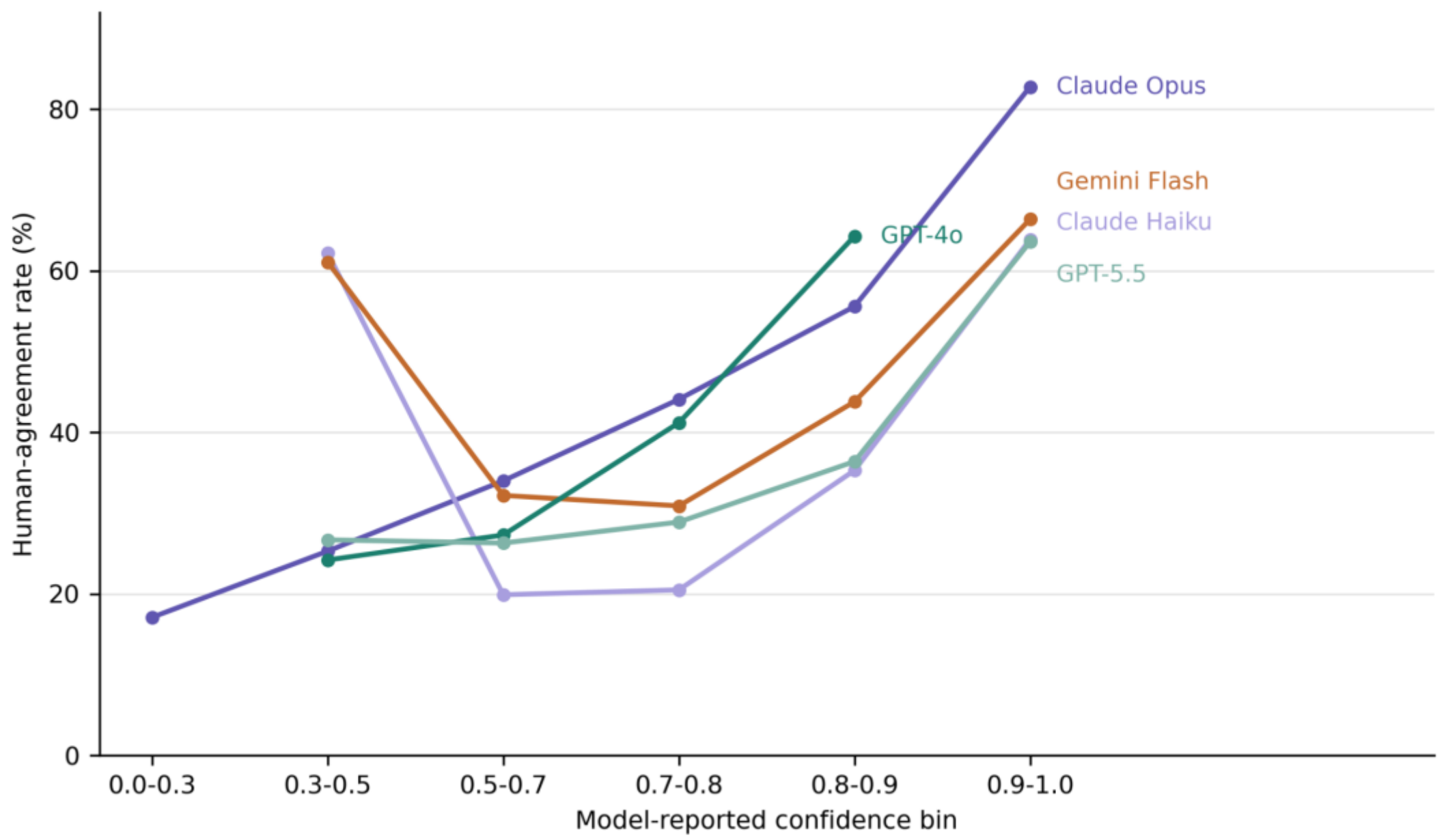

*Note. Identical bins for all models; bins with fewer than 30 annotations omitted. Claude Opus rises monotonically across its range, GPT-4o emits no confidence above 0.9 so its line ends at the 0.8 to 0.9 bin, and Gemini Flash discriminates only in its top bin.*

**RQ3: Blind Verification Findings**

*Overall Preference Direction*

Of 855 comparisons, the verifier expressed a decisive preference in 801 cases (93.7%), judged both sets adequate in 27 (3.2%), and judged neither adequate in 27 (3.2%). Among the 555 Human vs. LLM pairs, 515 received decisive preferences, with human coding preferred in 265 cases (51.5%) and LLM coding in 250 (48.5%). A binomial test yields $p = 0.537$, the 95% confidence interval for the human preference rate is [0.470, 0.559], and Cohen's h relative to 0.50 is 0.029, a negligible effect.

This finding directly addresses the central question of whether human coding is systematically preferred when disagreements occur. For the five models selected into verification, the answer is no. An independent domain expert, evaluating code sets blind to source, was equally likely to prefer the human or the LLM interpretation.

*Per-Model Preferences*

The aggregate null result masks substantial variation across models. Table 4 below presents the LLM win rate against human coders for each model.

**Table 4.**

*Per-Model LLM Win Rate Against Human Coders*

| Model | Decisive H-L pairs | LLM win rate | 95% CI | p-value |
|---|---|---|---|---|
| **Claude Opus** | 102 | 58.8% | [0.486, 0.685] | 0.092 |
| **Gemini Flash** | 102 | 57.8% | [0.477, 0.676] | 0.137 |
| **GPT-5.5** | 106 | 49.1% | [0.392, 0.590] | 0.923 |
| **GPT-4o** | 103 | 39.8% | [0.303, 0.499] | 0.048 |
| **Claude Haiku** | 102 | 37.3% | [0.279, 0.474] | 0.013 |

*Note.* *For each LLM's decisive Human vs. LLM pairs in the blind verification protocol, the percentage of comparisons in which the LLM's coding was preferred over human coding, with 95% confidence intervals and binomial test p-values against a null of 50%.*

Claude Opus and Gemini Flash were preferred over human coders more often than not, though neither difference reaches per-model significance. GPT-5.5 shows no directional preference. GPT-4o and Claude Haiku were preferred significantly less often than human coders ($p = 0.048$ and $p = 0.013$), confirming lower coding quality as judged by the verifier.

*Bradley-Terry Rankings*

A Bradley-Terry model fit to all 801 decisive comparisons produces the following ranking of the eight verified sources.

**Figure 3**.

*Bradley-Terry Quality Ranking of the Eight Verified Coding Sources*

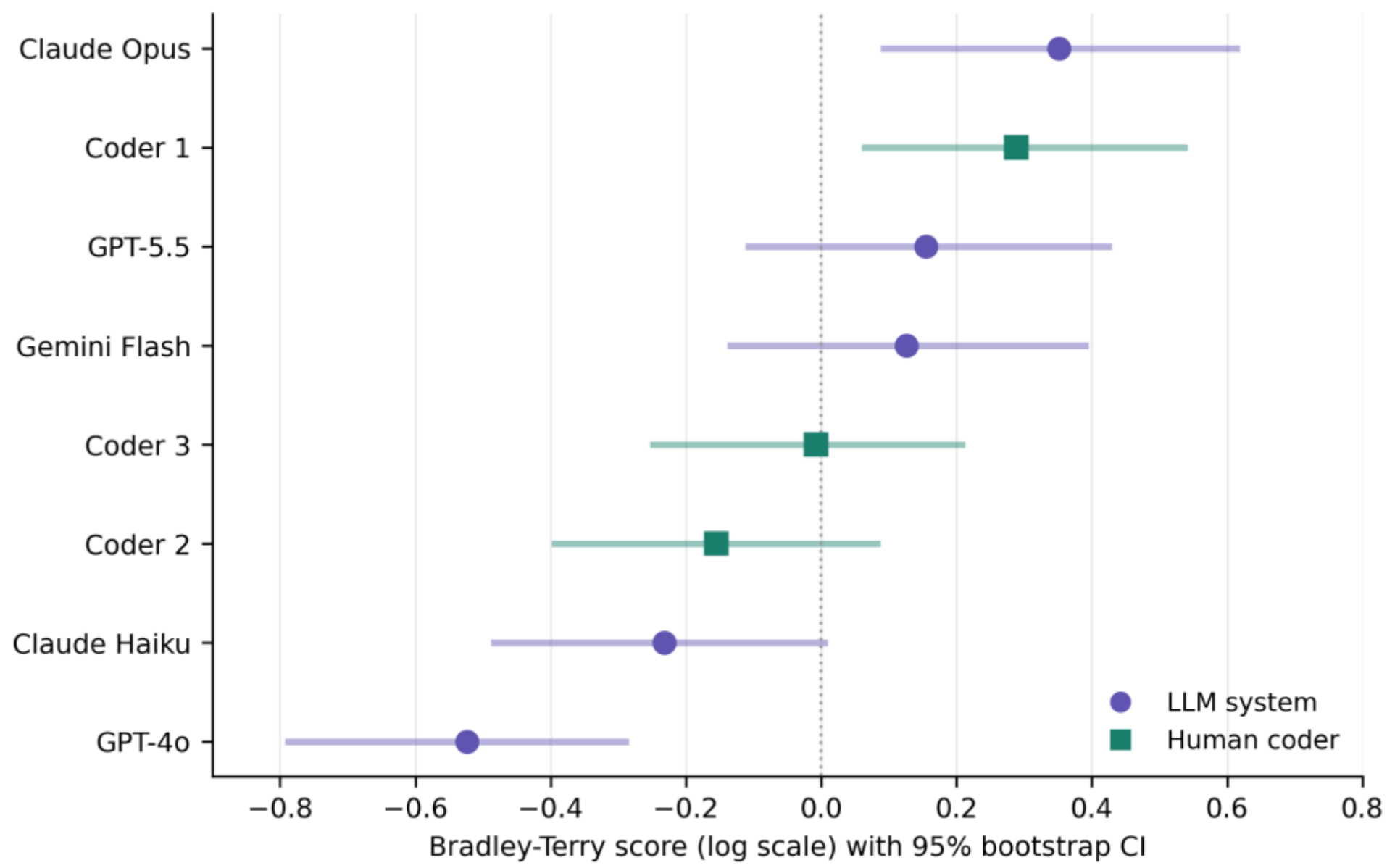

*Note.* *Estimated from 801 decisive blind pairwise comparisons. Points show log-scale ability estimates and horizontal bars show 95% bootstrap confidence intervals, with human coders and LLM systems distinguished by marker style. The interleaved ordering shows that source type*

*does not determine quality, while the 0.875 log-unit spread among LLMs dwarfs the 0.063 log-unit gap between the best LLM and the best human coder.*

Claude Opus occupies the top position with a confidence interval overlapping substantially with Coder 1, so the two are statistically indistinguishable. GPT-5.5 and Gemini Flash rank above two of the three human coders, Claude Haiku ranks below all of them, and GPT-4o is the only source whose interval does not overlap with any human coder. Figure 3 presents the ranking graphically, showing both the interleaving of human and machine sources and the far larger spread among models than between the best model and the best human. BT scores and 95% bootstrap confidence interval for each source are in Appendix Table E2.

*Position Bias and Baseline Checks*

No significant position bias was detected (Option A 47.8% vs. Option B 52.2% of decisive cases, $p = 0.230$). Among Human vs. Human pairs, decisive preferences split nearly evenly across the three coders (53.6%, 48.2%, 48.1%), confirming that the task admits meaningful discrimination without favoring any individual coder.

*Per-Code Endorsement and the Divergence Between Agreement and Quality*

For codes appearing in only one of the two sets (contested codes), we computed how often the code's presence was endorsed versus how often its absence was preferred. Per-code endorsement rates rest on modest numbers of contested pairs (median n of 18), so individual percentages should be read as indicative, and the four patterns below are defined by consistent direction rather than precise magnitude. Across 74 codes with endorsement data, four patterns emerge when endorsement is triangulated with the agreement metrics from RQ1 and RQ2; the full per-code data appear in Appendix D.

**Pattern 1: High agreement confirms automation viability.** Codes such as Special Education (IEP) show both high human-LLM agreement (Jaccard 0.53) and comparable endorsement rates (human 64%, LLM 64%). Agreement and verification converge, supporting automation.

**Pattern 2: High human-human agreement, but the verifier prefers LLM coding.** Several substantive codes show moderate-to-high human-human agreement yet the verifier more often endorses the LLM interpretation. ELA Skills Development (H-H = 0.48) is endorsed 61% of the time when an LLM applies it but 30% when a human applies it, and Entire Lesson Planning (H-H = 0.50) shows the same pattern (56% vs. 32%). Unit Planning (H-H = 0.56) is endorsed in 79% of LLM applications while none of its three contested human applications were endorsed, a suggestive figure given the small count. Human consensus can thus reflect a shared conservative interpretation that an independent expert considers inappropriately narrow; two coders can reliably agree while both under-applying a code.

**Pattern 3: Low agreement, but LLM coding is preferred.** Inquiry and Deep Questions (H-L = 0.15) shows human endorsement of 23% versus LLM endorsement of 65%. The low agreement reflects LLMs applying the code where humans do not, yet the verifier judges the LLM application appropriate far more often than the human omission. This pattern appears across 12 codes.

**Pattern 4: Low agreement, and human coding is strongly preferred.** Tiered Scaffolding (human endorsed 86%, LLM 27%), Generate Summative Assessments (75% vs. 21%), and Multimedia Use for Instruction (71% vs. 35%) represent codes where LLMs over-apply and the verifier rejects their application. These codes require recognizing pedagogical intent not explicitly stated in the text.

The implication of Patterns 2 and 3 is that agreement alone cannot determine automatability. Figure 4 maps every code with complete data onto the two dimensions (full underlying data is available in Appendix D table).

**Figure 4**.

*Divergence between Agreement and Expert Endorsement at the Code Level*

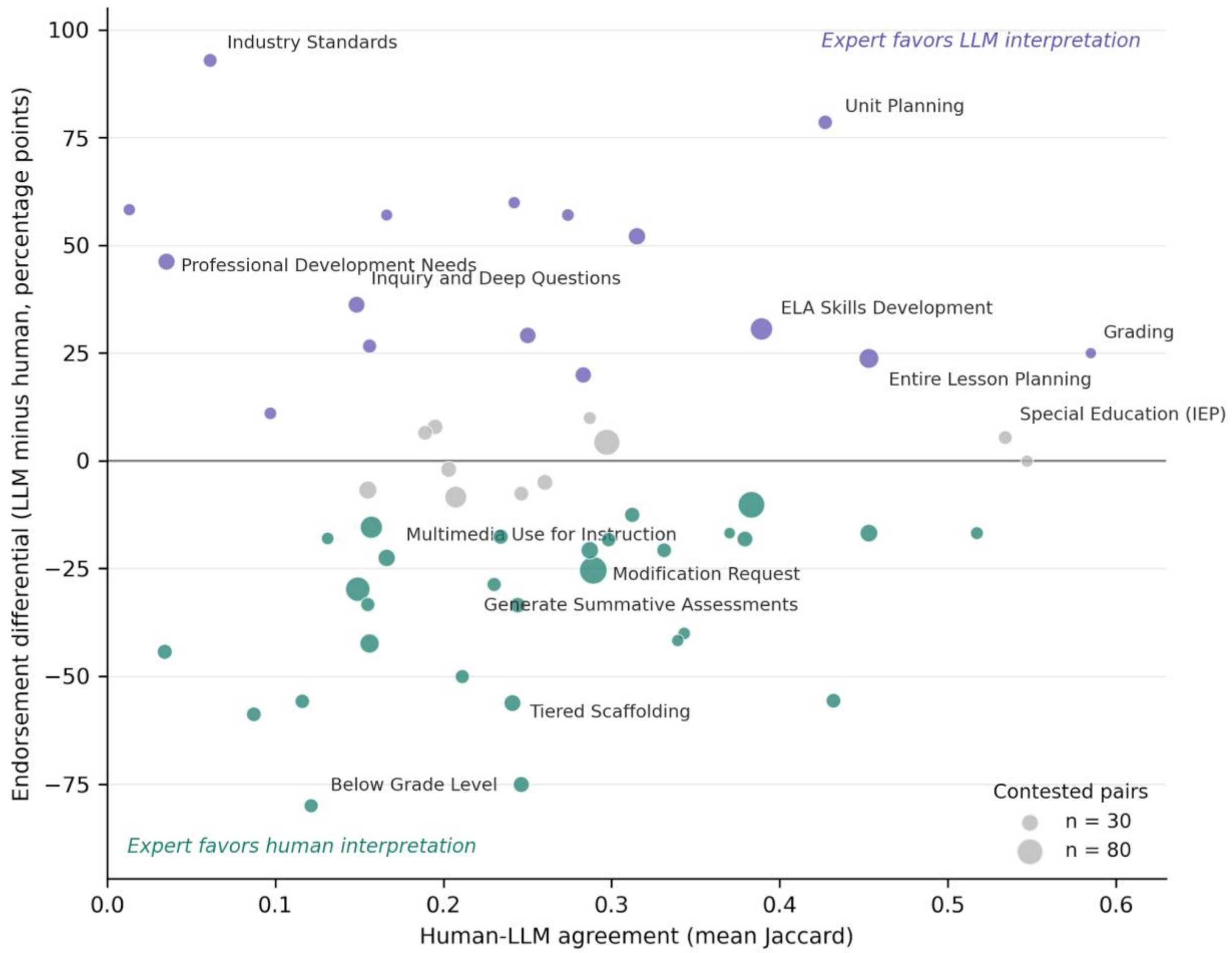

*Note. Each point is one code with complete data (57 codes); the horizontal axis is human-LLM agreement (mean Jaccard), the vertical axis is the endorsement differential (LLM minus human endorsement on contested applications), and point area is proportional to contested pairs. If agreement predicted quality, points would form a gradient; instead they populate all four*

*regions. Labeled codes mark the four patterns in the text; percentages for small-count codes, including Unit Planning, are indicative.*

**RQ4: Division of Labor**

Triangulating agreement with verification endorsement shows that the division of labor cannot be determined from agreement alone; it requires a two-dimensional classification considering both how often sources converge and which source an expert prefers when they diverge. This yields four categories, with full code lists in Appendix B.

**Automatable (6 codes).** These codes combine high human-LLM agreement (Jaccard above 0.40) with comparable or higher LLM endorsement. Four codes meet both criteria fully (Grading, Special Education (IEP), Entire Lesson Planning, Unit Planning) and two are automatable with light human review (Learning Standards Alignment, English Language Learners). For these codes, LLMs can serve as primary coders with minimal oversight.

**LLM-assisted (16 codes).** These codes show moderate agreement (Jaccard 0.20 to 0.40) and adequate LLM endorsement, particularly at high confidence. For models with calibrated confidence, accepting high-confidence annotations and routing low-confidence cases to human review is an effective triage; the confidence lift for these codes averages 40 percentage points.

**LLM-preferred (12 codes).** These codes show low human-LLM agreement, but the verifier consistently endorses LLM application more than human application, for example Inquiry and Deep Questions and Research and Source Analysis. Replacing human coding with LLM coding here would improve rather than degrade quality.

**Human required (15 codes).** These codes show human endorsement two to five times higher than LLM endorsement, for example Tiered Scaffolding (86% vs. 27%) and Generate

Summative Assessments (75% vs. 21%). They require recognizing pedagogical intent from context, a capability current models do not reliably demonstrate.

# Discussion

## Agreement Metrics and Quality Judgments Can Diverge

The central finding confirms the concern that motivated the study's design. Agreement metrics and normative quality judgments diverge in both directions. The agreement analysis alone would suggest that LLMs perform worse than humans (Jaccard 0.30 vs. 0.52), yet the blind verification reveals no overall expert preference for human coding (51.5% vs. 48.5%, $p = 0.537$). These findings are not contradictory; they expose a gap between convergence (how often sources agree) and quality (which interpretation is better when they disagree).

The gap manifests at the code level in ways that challenge standard assumptions. For ELA Skills Development, human coders agree with one another at Jaccard 0.48, which standard practice would read as evidence of a shared valid framework, yet the blind verifier endorses the LLM interpretation twice as often (61% vs. 30%). Entire Lesson Planning and Unit Planning show the same pattern, the latter on a small contested sample. Human coders agreed on a conservative application of these codes, and an independent expert judged the shared conservatism as under-coding. This provides empirical support for the shift from agreement-as-quality to disagreement-as-information proposed by Thomas et al. (2025). The implication is that agreement-only evaluation, however sophisticated the metric, is insufficient for determining the role of LLMs in qualitative coding, and triangulation with independent quality assessment is necessary.

## Model Selection Matters More Than Source Type

While the aggregate result shows no human-LLM preference, the per-model results show that the choice of model matters more than the choice between human and machine. The Bradley-Terry model places the best LLM (Claude Opus, +0.351) above the best human coder (Coder 1, +0.288), while the weakest model (GPT-4o, -0.524) ranks significantly below all human coders; the distance between the best and worst LLM exceeds the distance between the best LLM and the best human by a factor of fourteen. A team that selects a model on cost or convenience without evaluating it on their instrument could produce substantially worse results than human coding, while a team that invests in model evaluation could match its best human coder. Model tier within a provider predicts outcomes (Opus substantially outperforms Haiku), though not strictly across providers, since Gemini Flash outranks the newer GPT-5.5.

**Confidence Calibration Is Model-Specific**

Model confidence provides a useful triage signal, but its informativeness is model-specific in both strength and shape. Claude Opus discriminates across its entire range, with agreement rising 54 points from pooled low-confidence to high-confidence annotations. GPT-4o also discriminates, but it never emits confidence above 0.9, so a triage threshold set at 0.9 would route every one of its annotations to human review. Gemini Flash concentrates over 91% of its annotations at 0.7 or above and is nearly flat below 0.9, so thresholds in that region carry little information for that model. Hybrid workflows that route uncertain annotations to human review are therefore viable only when a model's calibration and emitted confidence range have been tested on the specific instrument before deployment.

**The Boundary of (Current) LLM Capability**

The per-code endorsement analysis identifies 15 codes where human expertise is demonstrably necessary, with endorsement ratios of two to five favoring human application.

These codes share a common property: they require recognizing pedagogical intent that is implied rather than stated. Tiered Scaffolding requires inferring differentiation by level, Generate Summative Assessments requires distinguishing summative from formative contexts, and Multimedia Use for Instruction requires recognizing instructional rather than administrative deployment of a tool. Until models develop richer pedagogical reasoning, or prompting makes this implicit knowledge explicit, these codes cannot be automated without quality loss.

The complementary finding is equally important. For 12 codes, the verifier consistently endorses LLM application over human application, for example Inquiry and Deep Questions (65% vs. 23%) and Research and Source Analysis (62% vs. 33%). LLMs detect relevant pedagogical content that trained coders systematically miss, whether through attentional limits during high-volume coding, conservative calibration, or genuine blind spots. For these codes, LLM coding is an improvement rather than a compromise.

**Implications for Practice**

Four guidelines follow for researchers designing large-scale coding pipelines. First, model evaluation on the target instrument should precede deployment, since the gap between the best and worst model is far larger than the gap between the best model and the best human. Second, the division of labor should be set at the code level using the two-dimensional assessment; blanket decisions in either direction are suboptimal, and the four-category framework provides a principled basis for allocation. Third, confidence-based triage should be used only after testing that high-confidence annotations are in fact endorsed more often than low-confidence ones for the chosen model. Fourth, high human-human agreement should not be treated as sufficient evidence that human coding is correct, since for codes where coders share a conservative interpretation, LLMs may produce more appropriate coding, and this pattern appears in substantive rather than marginal categories.

### Situating Within the Literature

The divergence finding speaks directly to concerns raised across the recent literature. It qualifies claims of model superiority such as Törnberg (2025), whose evaluation relied on a setting where ground truth was available; the present results show that where no ground truth exists, superiority claims in either direction require source-symmetric evidence. Ashwin et al. (2025) demonstrated that LLMs introduce systematic bias; the present study shows that human coders do as well, in the form of conservative under-coding on certain categories. Baumann et al. (2025) warned that LLMs can produce consistent but biased outputs that agreement metrics score as reliable; the present data confirm the concern and add its converse, since low agreement does not guarantee that the LLM interpretation is worse. The finding that LLMs are preferred for 12 codes extends the complementary-strengths framing of Liu and Sun (2025) from the descriptive claim that humans and LLMs excel on different code types to the normative claim that, for certain codes, the LLM interpretation is better as judged by an independent expert.

The evaluation setting strengthens the external validity of the paradigm argument. The 72-code instrument applied to unstructured naturalistic data is more demanding than most prior evaluations, and these are the conditions under which the gold standard assumption is weakest. That the best LLMs match the best human coders even here suggests findings from simpler instruments (Dunivin, 2025; Chew et al., 2023) extend to ecologically valid settings, and it establishes that the divergence documented here arises under realistic research conditions rather than as an artifact of a contrived task

## Limitations

Several limitations present in the study. First, the study employs a single codebook on a single dataset, so the findings, and especially the specific codes identified as automatable or human-required, may not generalize to other domains, data structures, or coding approaches. Second, the verification relies on a single independent verifier. A panel would provide stronger normative evidence and allow assessment of inter-verifier reliability, and a different expert might endorse the human coders' interpretation, so the divergence findings demonstrate that divergence is possible rather than that human coding is systematically wrong. Third, the LLMs represent a snapshot of capability; specific rankings will shift with successive generations, while the structural findings are more likely to persist. Fourth, all inference used temperature zero with no session memory, which maximizes reproducibility but leaves enhanced prompting strategies unexplored. Fifth, the models entering verification were a subset of those that coded the corpus, selected with attention to capability and adoption under a fixed verification budget, so the aggregate finding of no human-LLM preference is conditional on this model mix, though the central divergence finding is observed within the verified set and does not depend on which models were excluded.

## Conclusion

This study set out to test the validation logic on which the field's automation decisions currently rest, namely that agreement with trained human coders is an adequate measure of LLM coding quality. By subjecting five LLM systems and three human coders to the same blind expert evaluation on a 72-item codebook applied to unstructured K-12 educator messages, the study shows that this logic fails in both directions. The blind verification reveals no overall preference for human over LLM coding (51.5% vs. 48.5%, $p = 0.537$), the Bradley-Terry model places the best LLM statistically level with the best human coder, and for several substantive codes human coders agree with one another while the verifier prefers the LLM interpretation, indicating that human consensus can reflect shared conservative bias rather than superior quality.

These findings argue against both blanket automation and blanket human-only approaches. Six codes are automatable, four fully and two with light review. Twelve codes are better served by LLM coding than by human coding, fifteen demonstrably require human expertise, and sixteen benefit from confidence-based triage with calibrated models. LLM-assisted coding is therefore viable for scaling deductive qualitative analysis when researchers invest in model evaluation, adopt code-specific delegation, and retain human oversight for codes requiring contextual inference. The stratified workflow and verification protocol demonstrate a transferable methodology for researchers employing alternative codebooks or distinct data corpora.

The methodological implication reaches further. When an independent expert is equally likely to prefer LLM coding as human coding, and when human consensus itself can diverge from expert judgment, agreement-based evaluation alone cannot ground automation decisions. As AI-assisted analysis becomes available across the social sciences, the credibility of the resulting evidence will depend less on whether models imitate human coders and more on whether the field develops validation practices capable of detecting when either source is better at answering the proposed questions. The blind, source-symmetric protocol demonstrated here offers one transferable step in that direction.